\documentclass[12pt,a4]{article}

\usepackage{amsthm,amsmath,latexsym,amssymb,amsfonts,amscd}
\usepackage{graphics,lscape,fancyhdr,array,stmaryrd,euscript}
\usepackage{color}
\usepackage{relsize}
\numberwithin{equation}{section}
\usepackage{hyperref,setspace}
\usepackage[stable]{footmisc}

\usepackage[numbers,sort&compress]{natbib}
\usepackage[nottoc]{tocbibind}

\newcommand{\pl}{\partial}

\newcommand{\be}{\begin{equation}}
\newcommand{\ee}{\end{equation}}

\newcommand{\fud}[2]{{}^{#1}{}_{#2}\,}
\newcommand{\fdu}[2]{{}_{#1}{}^{#2}\,}

\newcommand{\besubeqs}{\begin{subequations}}
\newcommand{\esubeqs}{\end{subequations}}

\newcommand{\Sp}{{\ensuremath{\mathsf{p}}}}

\newcommand{\aI}{{\ensuremath{\mathcal{I}}}}
\newcommand{\aJ}{{\ensuremath{\mathcal{J}}}}

\newcommand{\aAb}{{\ensuremath{\boldsymbol{\mathcal{A}}}}}
\newcommand{\aBb}{{\ensuremath{\boldsymbol{\mathcal{B}}}}}

\newcommand{\JJJ}{{\boldsymbol{J}}}
\newcommand{\HHH}{{\boldsymbol{H}}}
\newcommand{\PPP}{{\boldsymbol{P}}}
\newcommand{\LLL}{{\boldsymbol{L}}}
\newcommand{\KKK}{{\boldsymbol{K}}}
\newcommand{\DDD}{{\boldsymbol{D}}}

\newcommand{\XXX}{{\boldsymbol{X}}}

\newcommand{\PP}{{\mathbb{P}}}

\newcommand{\breb}{\breve{\beta}}
\newcommand{\delbeta}{\Delta_{\beta}}
\newcommand{\MM}{\mathbb{M}}
\newcommand{\MMc}{\mathsf{M}}
\newcommand{\Nb}{\mathbb{N}}

\newcommand{\Tr}{\text{Tr}}
\newcommand{\Db}{{\Delta_\beta}}

\newcommand{\JJ}{{\mathbb{J}}}

\newcommand{\VV}{{\mathbf{V}}}

\begin{document}

\pagenumbering{gobble}
\hfill
\vskip 0.01\textheight
\begin{center}

{\Large\bfseries 
Light-Front Bootstrap \\
\vspace{0.4cm}
for Chern-Simons Matter Theories}

\vskip 0.03\textheight

Evgeny Skvortsov${}^{a,b}$ 

\vskip 0.03\textheight

{\em$^a$  Albert Einstein Institute, \\
Am M\"{u}hlenberg 1, D-14476, Potsdam-Golm, Germany}
\vspace{5pt}

{\em$^b$ Lebedev Institute of Physics, \\
Leninsky ave. 53, 119991 Moscow, Russia}\\

\end{center}

\vskip 0.02\textheight

\begin{abstract} 
We propose a new approach to solve conformal field theories and apply it to Chern-Simons Matter theories and three-dimensional bosonization duality. All three-point correlation functions of single-trace operators are obtained in the large-$N$ as a simple application. The idea is to construct, as an effective weakly-coupled theory, a nonlinear realization of the conformal algebra in terms of physical, gauge-invariant, operators. The efficiency of the method is also in the use of an analog of the light-cone gauge and of the momentum-space on the CFT side. AdS/CFT is used as a convenient regulator and as a source of the canonical bracket. The uniqueness of the nonlinear realization manifests the three-dimensional bosonization duality at this order. We also find two more non-unitary solutions which should be analogous to the fishnet theories. The results can also be viewed as an explicit realization of the slightly-broken higher spin symmetry.

As a by-product, the cubic action of the Higher Spin Gravity in $AdS_4$ is constructed. While generic Higher Spin Gravities are obstructed at higher orders by nonlocality, we point out the existence of two especially simple and well-defined theories: chiral and anti-chiral whose three-point functions correspond to the two new solutions. These two theories are supposed to give a quantum complete and local example of gravitational bulk duals.
\end{abstract}

\newpage
\section*{Introduction and Main Results}
\pagenumbering{arabic}
\setcounter{page}{2}
In the paper we propose a new approach to bootstrap nontrivial conformal field theories. It is based on the light-front approach to field and string theories \cite{Goddard:1973qh,Bengtsson:1983pg,Metsaev:1991nb,Brink:2005wh,Metsaev:2018xip} and, in few words, allows one to construct a nonlinear realization of the conformal algebra in terms of primary operators. Our main interest are three-dimensional conformal field theories of $O(N)$ vector-model type and, in particular, Chern-Simons Matter theories, see e.g. \cite{Giombi:2011kc, Maldacena:2012sf, Aharony:2012nh,GurAri:2012is}, and closely related ones. These CFT's have recently been conjectured to exhibit a number of remarkable dualities \cite{Giombi:2011kc, Maldacena:2012sf, Aharony:2012nh,Aharony:2015mjs,Karch:2016sxi,Seiberg:2016gmd}, including the three-dimensional bosonization duality. As an application, we show that the light-front bootstrap immediately gives all three-point correlation functions of single-trace operators in Chern-Simons Matter theories. This manifests the three-dimensional bosonization duality to this order by showing the solution for the correlators to be unique up to an expected number of phenomenological parameters. The nonlinear realization of the conformal algebra provides thereby an explicit algebraic implementation of the slightly-broken higher spin symmetry \cite{Maldacena:2012sf}. In addition we also find two more new solutions, which should be $3d$ analogs of the fishnet theories \cite{Gurdogan:2015csr,Caetano:2016ydc}. 

The $AdS/CFT$ correspondence for CFT's with matter in vectorial representations is very much different from the usual $AdS/CFT$ \cite{Maldacena:1997re,Witten:1998qj,Gubser:1998bc}  in that the spectrum of single-trace operators is not sparse and the gravitational dual description should be given by hypothetical Higher Spin Gravities.  The case of $AdS_4/CFT_3$ vectorial duality \cite{Klebanov:2002ja,Sezgin:2002rt,Sezgin:2003pt,Leigh:2003gk,Giombi:2011kc} is even more special in that the vector models can be generalized as to form families of CFT's, the Chern-Simons Matter theories. There are four most basic theories of this type, obtained by coupling free/critical boson/fermion CFT's to Chern-Simons action. For these theories, at least in the large-$N$ limit, the single-trace operator spectrum includes an infinite number of (almost) conserved tensors. In the free limit these are the old 'zilch' currents of type $J_s=\phi \pl...\pl \phi+...$, or the higher spin currents in modern terminology. It is clear that free CFT's have infinite dimensional symmetries. The algebra of symmetries is generated by the higher spin currents. Any nontrivial interaction breaks the conservation of these currents, see e.g. \cite{Deser:1980fk,Maldacena:2011jn,Boulanger:2013zza,Alba:2013yda,Alba:2015upa} for various contexts. Therefore, in general the infinite-dimensional symmetries of the free limits are broken by interactions and may be of little use.

In vector models the breaking of the conservation turns out to be of a very special type, the {\it non-conservation} operator is built out of the higher spin currents themselves:
\begin{align}\label{intro}
    \pl \cdot J&= g\, [JJ] \tag{$\star$}\,,
\end{align}
where the coupling $g$ is of order $1/N$. 
The non-conservation is an exact quantum non-perturbative equation of motion, in principle. It describes how a short multiplet of higher spin currents recombines with a long multiplet of double-trace operators $[JJ]$ as to form an even longer multiplet of anomalous currents. In practice, it is not easy to extract the CFT data. One approach is to directly use one or another (thanks to the dualities) microscopical realization of these CFT's via Chern-Simons Matter theories. However, even the leading order computations are quite challenging \cite{Giombi:2011kc,Aharony:2012nh,GurAri:2012is,Aharony:2012ns,Jain:2013py}, which is in contrast with the simplicity of the physical observables. Despite the significant progress in formulating and checking the dualities, even the three-point functions of the single-trace operators are not fully known in the large-$N$ limit \cite{Maldacena:2012sf,Giombi:2016zwa}, which is the gap we bridge in the paper.

Another approach\footnote{We discuss analytical approaches, but it would be very interesting to extend the numerical bootstrap \cite{ElShowk:2012ht} to Chern-Simons Matter theories.} is to investigate the non-conservation equation \eqref{intro}, the main advantage being in that we avoid working with the fundamental constituents, bosons or fermions, that are not observable. At the very least it can be used to fix the form of the three-point correlation functions and to extract the leading anomalous dimensions of the higher spin currents \cite{Maldacena:2012sf,Giombi:2016hkj,Skvortsov:2015pea,Giombi:2016zwa,Aharony:2018npf,Alday:2016jfr,Charan:2017jyc,Yacoby:2018yvy,Sleight:2018ryu,Turiaci:2018dht}. Algebraically, the non-conservation equation determines the deformation of the free CFT's infinite dimensional symmetry, which is called {\it slightly-broken higher spin symmetry} \cite{Maldacena:2012sf}. We expect that the slightly-broken higher spin symmetry should be powerful enough as to fix all correlation functions thereby proving the bosonization duality at least in the large-$N$ limit. 

In order to solve the models we suggest to isolate the subsector of single-trace operators and attempt to construct an effective weakly-coupled theory that computes the correlation functions in this subsector. The only requirement is that there should exist operator-dependent generators that form the conformal algebra. In the free approximation the generators are quadratic in the single-trace operators. Three-point correlators amount to cubic corrections to the generators and so on. Therefore, we simply construct a nonlinear realization of the conformal algebra.
This nonlinear realization is a direct implementation of the idea of the slightly-broken higher spin symmetry, see also \cite{Sharapov:2018kjz} for a more abstract approach.

The AdS/CFT correspondence \cite{Maldacena:1997re,Gubser:1998bc,Witten:1998qj} offers one more advantage --- the conformal algebra can be realized by bulk fields for which there exists a canonical Poisson bracket. Therefore, when extended to the bulk, the analysis can also be viewed as bootstrap of Higher Spin Gravities. These theories are defined as duals of the vector models and Chern-Simons Matter theories \cite{Klebanov:2002ja,Sezgin:2002rt,Sezgin:2003pt,Leigh:2003gk,Giombi:2011kc}. It is worth stressing that the bulk interpretation of the results is an additional bonus, but the whole procedure of constructing a nonlinear realization can be implemented directly on the CFT side. This is important in view of the fact that higher spin theories do not seem to exist as field theories due to severe non-localities \cite{Bekaert:2015tva,Sleight:2017pcz,Ponomarev:2017qab}, which is an immediate consequence of the degeneracy of the single-trace operators' spectrum taking place in vector-models.\footnote{There is no problem to 'simply' reconstruct any AdS dual, e.g. Higher Spin Gravities, from the CFT correlation functions \cite{Heemskerk:2009pn,Bekaert:2015tva,Sleight:2016dba}, provided that the correlators are already known. For the vector models a non-perturbative reconstruction seems possible \cite{Koch:2010cy}.}

Concerning the three-dimensional bosonization conjecture, our approach is insensitive to the actual microscopical constituents of the higher spin currents. Therefore, we investigate the consequence of the slightly-broken higher spin symmetry \cite{Maldacena:2012sf}.  We find the expected solution for three-point correlation functions with two phenomenological parameters, $\tilde N$ and $\theta$, which are related to the microscopical number of fields $N$ and Chern-Simons level $k$:
\begin{align}\notag
    \langle J_{s_1}J_{s_2}J_{s_3}\rangle&=\tilde{N} \left(\cos^2\theta \langle J_{s_1}J_{s_2}J_{s_3}\rangle_{F.B.}+\sin\theta\cos\theta\langle J_{s_1}J_{s_2}J_{s_3}\rangle_{Odd}+\sin^2\theta \langle J_{s_1}J_{s_2}J_{s_3}\rangle_{F.F}\right)\,,
\end{align}
where $F.B.$ means as computed in the free boson CFT, $F.F.$ the free fermion CFT and Odd is one more structure specific to three dimensions \cite{Giombi:2011rz}. This simple structure is in contrast with the technical difficulties of computations in Chern-Simons Matter Theories 
\cite{Giombi:2011kc,Aharony:2012nh,GurAri:2012is,Aharony:2012ns,Jain:2013py}. One of our goals is to extend \cite{Maldacena:2012sf} to all spins and to give explicit expressions for all the structures, in particular for the odd one. The even ones, which arise from free CFT's, can also be understood as invariants of the unbroken higher spin symmetry  \cite{Colombo:2012jx,Didenko:2012tv,Didenko:2013bj,Sleight:2016dba, Bonezzi:2017vha}. We also find two new isolated solutions for the three-point functions, which are non-unitary and should be close to the fishnet theory \cite{Gurdogan:2015csr,Caetano:2016ydc}, see also the discussion in section \ref{sec:corrandbos}.

Technically, an analog of the light-cone gauge that we employ on the CFT side reduces conserved tensors $J_s$ to just two scalars that are helicity eigen-states. This considerably simplifies computations and also removes the redundancy present in longitudinal components of conserved currents. The light-cone gauge would also be especially favorable in the AdS dual description of Chern-Simons Matter theories. The conservation of higher spin currents is dual to the gauge symmetry of bulk fields, which, of course, is just a redundancy of description. For this reason it is natural to remove the redundancy by working directly with the physical degrees of freedom, the two bulk helicity eigen states. The latter are dual to particular two conformal operators that solve the conservation law, the ones we use to construct the nonlinear realization of the conformal algebra. The AdS dual description can be thought of as a particular way to achieve the same goal of constructing the nonlinear realization.  

The light-cone gauge has always been very close to the spinor-helicity formalism in flat space. In particular, there is a one-to-one correspondence between three-point spinor-helicity amplitudes \cite{Benincasa:2007xk,Conde:2016izb} and cubic vertices in the light-cone gauge \cite{Bengtsson:1983pd,Metsaev:1991nb}. As it has been just shown \cite{Metsaev:2018xip}, each of the flat-space cubic vertices has a unique $AdS_4$ uplift. It is this uplift that we use to compute the three-point functions. Therefore, our work also establishes a dictionary between flat-space spinor-helicity amplitudes, cubic vertices in $AdS_4$ and correlation functions of conserved tensors in $CFT_3$, see also \cite{Maldacena:2011nz,Nagaraj:2018nxq}. It turns out that there are slightly more independent three-point structures that was previously seen via manifestly covariant methods \cite{Giombi:2011kc}.

Since we bootstrap Chern-Simons Matter theories with the help of the bulk representation of the correlation functions, we are able to fix the cubic action of their Higher Spin Gravity duals. The two new non-unitary solutions correspond to the (anti)-chiral Higher Spin Gravities. The latter are the $AdS_4$-uplifts of the Chiral Higher Spin Gravities discovered recently in flat space \cite{Metsaev:1991nb,Metsaev:1991mt,Ponomarev:2016lrm,Ponomarev:2017nrr,Skvortsov:2018jea}.  

Lastly, let us point out that there are at least three seemingly different realizations of the slightly-broken higher spin symmetry. The first one is via the non-conservation equation, broken Ward identities and it leads to \cite{Maldacena:2012sf}. The second one is via $A_\infty$-algebra constructed in \cite{Sharapov:2018kjz} that can be viewed as a deformation of the infinite-dimensional symmetries associated with higher spin currents $J_s$. The $A_\infty$-algebra of \cite{Sharapov:2018kjz} is closely related to the deformation quantization and to the formality theorem \cite{Kontsevich:1997vb} and, thereby, to topological string theory. The third one, in the present paper, is via non-linear realization of the conformal algebra in terms of the same higher spin currents in the light-cone gauge. While the equivariant map from the second, more covariant, realization to the third one is yet to be understood, it is already clear, as our results show, that they are consistent with each other in the sense of having the same number of free parameters and featuring the same special limits.

\noindent To summarize, our results are as follows. 
\begin{itemize}

    \item Using the light-front version of $AdS_4/CFT_3$-correspondence all possible structures for three-point correlation functions of conserved tensors $\langle J_{s_1} J_{s_2} J_{s_3}\rangle$ are obtained in momentum space. They are in one-to-one correspondence with the spinor-helicity amplitudes of three massless fields in $4d$ flat space and, hence, there are slightly more of them then it has been previously observed, e.g. in \cite{Giombi:2011rz,Maldacena:2012sf};
    
    \item We bootstrap the three-point functions in Chern-Simons Matter theories by constructing the nonlinear realization of the conformal algebra $so(3,2)$ in terms of the single-trace primaries, i.e. higher spin currents;
    
    \item The uniqueness of the three-point functions manifests the three-dimensional bosonization duality to this order and can be viewed as an explicit realization of the slightly-broken higher spin symmetry. Interestingly, the parity breaking parameter $\theta$ results from electromagnetic duality transformations in the bulk;
    
    \item As compared to the slightly-broken higher spin symmetry analysis of \cite{Maldacena:2012sf}, we find two additional (non-unitary) solutions;
    
    \item The conformal algebra generators are constructed via $AdS_4$ bulk realization, which gives the complete cubic action of the Higher Spin Gravity dual. In particular the two additional solutions correspond to the $AdS_4$ uplift of the Chiral Higher Spin Gravity \cite{Metsaev:1991nb,Metsaev:1991mt,Ponomarev:2016lrm,Ponomarev:2017nrr,Skvortsov:2018jea};
\end{itemize}

The outline of the paper is as follows. The light-front approach to massless fields in $AdS_4$ is discussed in section \ref{sec:LCAdS} together with the review of the important recent results of \cite{Metsaev:2018xip}. The three-point functions in Chern-Simons Matter theories are bootstrapped in section \ref{sec:bootstrap} together with the dual Higher Spin Gravities.

\section{Light-Cone Gauge in AdS}
\label{sec:LCAdS}
Any local description of a QFT in flat space can be understood as a convenient way to construct the conserved charges of the Poincare algebra in terms of classical and, then, quantum operators. This leads to a Poincare invariant $S$-matrix, under certain assumptions involving locality. Unless we attempt to directly bootstrap the $S$-matrix, the light-front approach offers a lot of flexibility, see e.g. \cite{Goddard:1973qh,Bengtsson:1983pg,Metsaev:1991nb,Brink:2005wh,Metsaev:2018xip}. The main advantage is in that the light-cone gauge is complete and there is no redundancy left that usually accompanies massless fields. Therefore, in the light-front approach one attempts to build the charges of the space-time symmetry algebra directly in terms of physical degrees of freedom. Nontrivial constraints on dynamics come from the requirement to maintain the algebra while the charges receive nonlinear corrections. The same idea can, of course, be applied to the cases where the relevant symmetry is conformal symmetry. There are two such situations: field theories in (anti)-de Sitter space and conformal field theories. Thanks to AdS/CFT duality these two are not completely independent. 

We review the light-cone approach to massless fields in $AdS_4$ \cite{Metsaev:1999ui,Metsaev:2013kaa} and, importantly, the results of  \cite{Metsaev:2018xip}, the main goal being to construct a base of all possible cubic interaction vertices (amplitudes) for massless fields of any spin. Here AdS/CFT is utilized as a convenient way to encode CFT correlation functions. Indeed, there is a one-to-one correspondence between the CFT and AdS kinematics. For example, any conformally-invariant $CFT_d$ three-point function $\langle O_1 O_2 O_3\rangle $ corresponds to a cubic vertex $V_3$ in $AdS_{d+1}$ such that the holographic correlation function obtained from $V_3$ is exactly $\langle O_1 O_2 O_3\rangle $. Other way round, the CFT three-point functions in momentum space admit a rather useful representation as triple $K$-integrals \cite{Bzowski:2013sza} that can be recognized as evaluation of the holographic correlation function. Another advantage of using the AdS language as to encode CFT correlation functions is that in the process of bootstrapping a CFT we learn about its $AdS$-dual. Lastly, the $AdS$-description of a CFT comes automatically equipped with the canonical Poisson bracket, which is then employed to construct the nonlinear realization of the conformal algebra in section \ref{sec:bootstrap}. The conformal (anti-de Sitter) algebra commutation relations are chosen to be
\besubeqs
\begin{align}
[\DDD,\PPP^a]&=-\PPP^a\,, & [\LLL^{ab},\PPP^c]&=\PPP^a\eta^{bc}-\PPP^b\eta^{ac}\,,\\
[\DDD,\KKK^a]&=+\KKK^a\,,  & [\LLL^{ab},\KKK^c]&=\KKK^a\eta^{bc}-\KKK^b\eta^{ac}\,,\\
[\PPP^a,\KKK^b]&=-\LLL^{ab}+\eta^{ab}\DDD\,, &[\LLL^{ab},\LLL^{cd}]&=\LLL^{ad}\eta^{bc}+\text{three more}\,.
\end{align}
\esubeqs

\subsection{Free Action}
On the CFT${}_3$ side we have conserved higher spin tensors $J_{a_1...a_s}$. They are dual to massless fields in $AdS_4$. In the covariant language, a massless spin-$s$ field is usually described by a Fronsdal field $\Phi_{\mu_1...\mu_s}$. The free equations of motion in $AdS_d$ with cosmological constant $\Lambda$ read:
\begin{align}\label{FronsdalEq}
\Box \Phi_{\mu_1...\mu_s}-\nabla_{\mu_1}\nabla^\nu\Phi_{\nu\mu_2...\mu_s}+\frac12\nabla_{\mu_1}\nabla_{\mu_2}\Phi^{\nu}{}_{\nu\mu_3..\mu_s}-M_s^2\Phi_{\mu_1...\mu_s}+2\Lambda g_{\mu_1\mu_2}\Phi\fdu{\mu_3...\mu_s\nu}{\nu}=0\,,
\end{align}
where symmetrization over the $\mu$ indices is implied. The mass-like term $M_s^2=-\Lambda((d+s-2)(s-2)-s)$ is fixed by gauge invariance under
\begin{align}
\delta\Phi_{\mu_1...\mu_s}&=\nabla_{\mu_1}\epsilon_{\mu_2...\mu_s}+\text{permutations}\,.\label{freesymm}
\end{align}
The Fronsdal field and the gauge parameter obey certain algebraic constraints:
\begin{align}
\Phi\fud{\nu\lambda}{\nu\lambda\mu_5...\mu_s}\equiv0\,,&& \epsilon\fud{\nu}{\nu\mu_4...\mu_s}\equiv0\,.
\end{align}
Now we would like to impose the light-cone gauge. It is convenient to choose the Poincare coordinates
\begin{equation}
    ds^2=\frac{1}{z^2}\,\eta_{\mu\nu}dx^\mu dx^\nu=\frac{1}{z^2}\Big[2dx^+dx^-+dx_1^2+dz^2\Big]\,.
\end{equation}
Each massless spin-$s>0$ field has two degrees of freedom both in flat and (anti)-de Sitter spaces. Therefore, without having to impose the light-cone gauge directly in the covariant formulation, it is clear that we will end up with two complex conjugate fields that are helicity eigen-states:
\begin{align}
    \Phi_{\pm s}(x,z)&\equiv\Phi_{\pm s}(x^+,x^-,x^1,z), &\left(\Phi_{+s}\right)^{\dagger}&=\Phi_{-s}\,.
\end{align}
It is convenient to Fourier-transform with respect to all coordinates but $z$:\footnote{As to avoid clumsy notation we sometimes use $\beta\equiv p^+$, $\rho\equiv p^1$ and $\gamma\equiv p^-$, i.e. $p=(p^+,p^-,p^1)\equiv (\beta,\gamma, \rho)$.}
\begin{align}
    \Phi(p;z)&=\frac{1}{(2\pi)^{3/2}}\int dx^-dx^1dx^+ \,e^{-ipx}\Phi(x^+,x^-,x^1;z)\,.
\end{align}
The conjugation rules for the Fourier transformed fields are 
\begin{align}
    \Phi^{\lambda}_{p,z}&\equiv \Phi^\lambda(p;z)\,, &(\Phi^{\lambda}_{p,z})^{\dagger}=\Phi^{-\lambda}_{-p,z}\,.
\end{align}
It is sometimes useful to assume that the bulk fields are matrix-valued, which corresponds to 'gauging' some Yang-Mills groups\footnote{In the most general case the dual CFT can have certain global symmetries, e.g. $U(M)$. In the gravitational dual description these global symmetries need to be gauged. } $\Phi^{\lambda}_{p,z}\equiv \Phi^\lambda(p;z)\fdu{\aI}{\aJ}$ and we will use the trace $\Tr[...]$ to make singlets. 
One can show, see e.g. \cite{Metsaev:1999ui,Metsaev:2013kaa}, that the free action for all massless fields looks like the one in flat space\footnote{Note that the helicity fields $\Phi_\lambda$ are related to certain  components of the Fronsdal field $\Phi_{\mu_1...\mu_s}$ rescaled by a factor of $z^{(d-1)/2}$. It is worth stressing that the miracle is thanks to masslessness, $AdS_4$ and light-cone gauge. In particular, it is not true in $AdS_{d+1}$, $d>3$. The real reason for the disappearance of the mass-like term and for having just a flat space $\square$ in the action is that the massless higher spin fields are conformally invariant in four dimensions. This is true for the on-shell states as representations of $so(3,2)$ that extend to $so(4,2)$ \cite{Mack:1969dg,Metsaev:1995jp}, but is not true for the gauge potentials $\Phi_{\mu_1...\mu_s}$ except for $s=0,1$. Therefore, the action in the light-cone gauge can easily be guessed to have $\Phi \square \Phi$-form thanks to conformal symmetry. \label{ft:action}}
\begin{align}
    S_2&=\frac12 \sum_\lambda \int dx^+dx^- dx^1dz\, \Tr[\Phi^\dag_\lambda(x;z)(2\pl_+\pl_-+\pl_1^2+\pl^2_z)\Phi_\lambda(x;z)]\,. \label{freeLCaction}
\end{align}
A small difference from the flat space is that we integrate over $z>0$. The momentum space analog is  
\begin{align}\label{freeAction}
    S_2&=\frac12 \sum_\lambda \int dz\,d^3p\, \Tr[\Phi^\dag_\lambda(p;z)(\pl^2_z-p^2)\Phi_\lambda(p;z)]\,.
\end{align}
The interaction vertices, in particular the cubic ones, are specified by their kernels
\begin{align*}
    \boldsymbol{V}_n&= \int d\Gamma_n^3\, \Tr\left[\prod_a \Phi_a^\dag\right] V(p_a; \partial_{z_a}; z)\, \delta_z\,,
\end{align*}
where the vertex depends on $n$ fields $\Phi^\dag_a\equiv \Phi^\dag_{\lambda_a}(p_a;z_a)$ and
\begin{align*}
    d\Gamma_n^3&\equiv\delta^3(\sum p_a) \prod d^3p_a\, dz_a\,dz\,, &
    \delta_z&\equiv\prod \delta(z-z_a)\,.
\end{align*}
The kernel $V$ can depend on all three-momenta $p_a$, radial coordinate $z$ and derivatives $\pl_{z_a}$ that, upon integration by parts, start acting on $\Phi_a$'s. An important feature of the light-cone gauge is that the vertices do not depend on $p^-$, i.e. they do not contain any explicit time dependence on the light-cone time $x^+$. 

\subsection{Conformal Algebra: Quadratic Generators}
So far we have free massless fields in $AdS_4$. One can construct the generators of the conformal, i.e. $AdS$-algebra. The canonical treatment of the free action \eqref{freeAction} with $x^+$ chosen to be the time coordinate leads to the following classical (Dirac) Poisson bracket ($\Sp\equiv (p^+,p^1)$):
\begin{equation}
    [\Phi^{\lambda}(\Sp,z,x^+),\Phi^{\mu}(\Sp',z',x^+)]=\delta^{\lambda,-\mu}\frac{\delta^2(\Sp+\Sp')\delta(z-z')}{2\beta}\,.
\end{equation}
This is an ingredient that we do not have a priori on the CFT side if we start off with the subsector of higher spin currents without assuming any microscopical realization. The free field realization of the $so(3,2)$ charges $Q_\xi$ is obtained in a canonical way
\begin{equation}
    Q_{\xi}=\int \beta\, d^2\Sp\, dz\, \Tr[\Phi^\dag_\lambda(p;z)O_{\xi}(x^+,\Sp,\partial_{\Sp},z,\partial_z)\Phi_{\lambda}(p;z)]\,,
\end{equation}
where the kernels $O_{\xi}$ for the kinematical generators are 
\besubeqs
\label{helicitybaseAdS}
\begin{align}
    \hat{P}^+&=\beta\,, \qquad\qquad\qquad\hat{P}^1=\rho\,, \\ \hat{J}^{+1}&=ix^+\rho+\pl_\rho\beta\,, \\
    \hat{J}^{+-}&=i x^+ P^-+\pl_\beta\beta\,,\\
    \hat{K}^+&=\frac{1}{2}\left(2ix^+\pl_\beta +z^2-\partial_{\rho}^2\right)\beta+ix^+ \hat D\,, \\ \hat{K}^1&=\frac{1}{2}\left(2ix^+\pl_\beta+z^2-\partial^2_{\rho}\right)\rho-\partial_{\rho}\hat{D}+\lambda z+ix^+\lambda\frac{\pl_z}{\beta}\,,\\
    \hat{D}&=ix^+\hat P^--\pl_\beta\beta - \pl_\rho \rho+z\partial_z+1\,,
\end{align}
\esubeqs
and for the dynamical are
\besubeqs
\begin{align}
    \hat P^-&=-\frac{\rho\rho}{2\beta}+\frac{\partial_z^2}{2\beta}\,,\\
    \hat{J}^{-1}&=\frac{\partial}{\partial \rho}P^--\rho\frac{\partial}{\partial \beta}-\lambda\frac{\partial_z}{\beta}\,,\\
    \hat{K}^-&=\frac{1}{2}(2ix^+ \pl_\beta+z^2-\partial^2_{\rho})P^--\partial_{\beta}\hat{D}+\frac{\lambda}{\beta}(\partial_z\partial_{\rho}-z\rho)-\frac{\lambda^2}{\beta}\,.
\end{align}
\esubeqs
It is worth noting here that the cosmological constant does not enter any of the free generators and will not show up in the interacting parts of the generators. Therefore, we will find only dimensionless parameters enter the nonlinear realization, which has to be the case for a CFT interpretation to make sense.

\subsection{Conformal Algebra: Cubic Deformations}
\label{sec:cubicdeformations}
The main result of \cite{Metsaev:2018xip} is the classification of three-point interaction vertices in $AdS_4$ and in the light-cone gauge. The $10$ generators of $so(3,2)$ can further be split into {\it kinematical}, which stay quadratic in the fields and are not deformed by interactions, and {\it dynamical}, which contain interactions via terms nonlinear in the fields. By choosing the Cauchy surface appropriately one can minimize the number of dynamical generators. It turns out that there are three dynamical generators in the light-cone gauge
\begin{align}
    \HHH\equiv\PPP^-\,, \qquad \JJJ\equiv \JJJ^{1-}\,,\qquad \KKK\equiv \KKK^{-}\,,
\end{align}
which is exactly like in four-dimensional flat space. The dynamical generators $\HHH, \JJJ,\KKK$ are corrected by interactions. A convenient basis to represent higher order corrections is\footnote{As always, one can study deformations at $x^+=0$ and then evolve the generators with the help of $\HHH$.}
\begin{align}\label{genseed}
    \boldsymbol{F}_n&= \int d\Gamma_n\, \Tr\left[\prod_a \Phi_a^\dag\right] F(\Sp_a; \partial_{z_a}; z)\, \delta_z\,,
\end{align}
where the generator $\boldsymbol{F}_n$ depends on $n$ fields $\Phi^\dag_a\equiv \Phi^\dag(p_a,z_a,\lambda_a)=\Phi(-p_a,z_a,-\lambda_a)$ and
\begin{align*}
    d\Gamma_n&\equiv\delta^2(\sum \Sp_a) \prod d^2\Sp_a\, dz_a\,dz\,, &
    \delta_z&\equiv\prod \delta(z-z_a)\,.
\end{align*}
Another type of a structure that also appears is
\begin{align*}
    \XXX\{\boldsymbol{F}_n\}&= \int d\Gamma_n\, \left(-\frac{1}{n} \sum \pl_{\rho_a}\Tr\left[\prod \Phi_a^\dag\right]\right) F'(\Sp_a; \partial_{z_a}; z)\, \delta_z\,.
\end{align*}
Interacting parts of the dynamical generators have the following form
\besubeqs
\begin{align}
    \HHH&=\sum_n \HHH_n\,,\\
    \JJJ&=\sum_n(\JJJ_n +\XXX \{\HHH_n\})\,, \\
    \KKK&=\sum_n(\KKK_n- \XXX \{\JJJ_n\} -\frac12 \XXX\{\XXX \{\HHH_n\}\})\,.
\end{align}
\esubeqs
Once the Hamiltonian $\HHH$ is known, the corresponding action reads
\begin{align}
    S&=S_2+\sum_{n>2}\boldsymbol{V}_n\,, &&\boldsymbol{V}_n=\int dx^+ \HHH_n\,.
\end{align}
At the first nontrivial, cubic, order the problem of constructing $\HHH$ reduces to analyzing various relations of type $[K,D_3]=D_3$ and $[D_2,D_3]=D_3$, where $K$ are the kinematical generators, $D_2$ and $D_3$ are the free and cubic parts of the dynamical ones. The analysis is rather involved. long chain of ingenious transformations and simplifications employed in \cite{Metsaev:2018xip} leads to the following types of vertices that can contribute to the Hamiltonian density given three fields with helicities $\lambda_{1,2,3}$:
\begin{align}\label{Ham}
    V^{\lambda_1,\lambda_2,\lambda_3}&= \left\{
	\begin{aligned}
			&z^{-1}\,, & &&& \lambda_i=0\,,\\
			&U_RV_R^0\,, & V_R^0&=\frac{(z\,\PP^R )^{\lambda_1+\lambda_2+\lambda_3}}{z\,\beta_1^{\lambda_1}\beta_2^{\lambda_2}\beta_3^{\lambda_3}}\,, && \Lambda>0\,,\\
			&U_LV_L^0\,, & V_L^0&=\frac{(z\,\PP^L )^{-\lambda_1-\lambda_2-\lambda_3}}{z\,\beta_1^{-\lambda_1}\beta_2^{-\lambda_2}\beta_3^{-\lambda_3}}\,, && \Lambda<0\,.
	\end{aligned}\right.
\end{align}
where $\Lambda=\lambda_1+\lambda_2+\lambda_3$ and the holomorphic momenta 
\begin{align}
    \PP^L&=\frac{1}{\sqrt{2}}(\PP+\PP_z)\,, &\PP^R&=\frac{1}{\sqrt{2}}(\PP-\PP_z)\,,
\end{align}
are built out of\footnote{It can be convenient to define $\breb_i=\beta_{i+1}-\beta_{i-1}$ modulo $3$, i.e. $\breve{\beta}_i=\{(\beta_2-\beta_3),  (\beta_3-\beta_1), (\beta_1-\beta_2)\}$.}
\besubeqs
\begin{align}
    \PP&=\frac13 \sum_a \breb_a p^1_a\,,&
    \PP_z&=\frac13 \sum_a \breb_a \pl_{z_a}\,.
\end{align}
\esubeqs
Eq. \eqref{Ham} is the complete base of all possible cubic deformations. 
Any linear combination of densities $V^{\lambda_1,\lambda_2,\lambda_3}$ can be used via \eqref{genseed} to generate corrections $\VV^{\lambda_1,\lambda_2,\lambda_3}$ to Hamiltonian $\HHH$. In the actual theories they should enter with certain coupling constants, $C^{\lambda_1,\lambda_2,\lambda_3}$, so that the full cubic deformation of the action is 
\begin{align}
    \VV_3=\sum_{\lambda_i}C^{\lambda_1,\lambda_2,\lambda_3} \VV^{\lambda_1,\lambda_2,\lambda_3}\,.
\end{align}
Our goal is to determine $C^{\lambda_1,\lambda_2,\lambda_3}$ that correspond to Chern-Simons Matter theories.   
The first term in \eqref{Ham} is the cubic vertex of the scalar field and $C^{0,0,0}$ is its coupling. The other two terms represent (anti)holomorphic vertices with their coupling constants $C^{\lambda_1,\lambda_2,\lambda_3}$. The helicities are constrained to be $\Lambda>0$ for $V_R$ and $\Lambda<0$ for $V_L$. Various combinations from $+++$ to $---$ are in principle allowed. In fact there is one-to-one relation between the light-cone vertices in $AdS_4$ and in flat space \cite{Bengtsson:1983pd,Metsaev:1991nb}, where the classification is identical to that in the spinor-helicity formalism \cite{Benincasa:2007xk}
\begin{align}
    V^{\lambda_1,\lambda_2,\lambda_3}_R\Big|_{\text{flat limit}}&\sim [12]^{\lambda_1+\lambda_2-\lambda_3}[23]^{\lambda_2+\lambda_3-\lambda_1}[13]^{\lambda_1+\lambda_3-\lambda_2}\,, &&\Lambda>0\,,\\
    V^{\lambda_1,\lambda_2,\lambda_3}_L\Big|_{\text{flat limit}}&\sim \langle 12\rangle^{-\lambda_1-\lambda_2+\lambda_3}\langle 23\rangle^{-\lambda_2-\lambda_3+\lambda_1}\langle 13\rangle^{-\lambda_1-\lambda_3+\lambda_2}\,,   &&\Lambda<0  \,.
\end{align}
Another important ingredient is the dressing map $U$:
{\begin{allowdisplaybreaks}
\besubeqs
\begin{align}
    U_{R,L}&=T\exp\Big[\int_0^1d \tau\,  u^{R,L}_\tau]\,,\\
    u^R_{\tau}&=\frac{1}{\sqrt{2}}(\Nb+\MM)Y_R-\frac{\tau \Db}{18}Y_R^2N_{\PP^R}\,, \\
    u^L_{\tau}&=\frac{1}{\sqrt{2}}(-\mathbb{N}+\MM)Y_L-\frac{\tau\Db}{18}Y_L^2N_{\PP^L}\,,\\
     Y_R&=\frac{1}{N_z+2}\partial_z\partial_{\PP^R}\,,\qquad Y_L=\frac{1}{N_z+2}\partial_z\partial_{\PP^L}\,,
\end{align}
\esubeqs
\end{allowdisplaybreaks}}
\noindent and a bit of new notation again
\besubeqs
\begin{align}
    \Nb&=\frac13 \sum_a \breb_a\beta_a\pl_{\beta_a}\,, &
    \MM&=\frac13 \sum_a \breb_a \lambda_a\,,
\end{align}
\esubeqs
and, for last, $N_z=z\pl_z$, etc., $\Db=\sum_i \beta_i^2$.

The most striking feature of the result \eqref{Ham} is that it looks exactly like the one in flat space up to the $U$-map and $z$-factors. Therefore, there is a one-to-one correspondence between vertices in flat space and $AdS_4$.   These vertices deform the dynamical generators of $so(3,2)$ algebra. From the conformal point of view, after the computation of the Witten diagram, they describe all possible three-point functions that are consistent with conformal symmetry. There is no yet any nontrivial input that would fix the coupling constants $C^{\lambda_1,\lambda_2,\lambda_3}$.

$U_{R,L}$ is a dressing map whose job is to complete the leading part of the vertex, $V^0$, with a tail of lower derivative terms that makes it consistent. This is easy to understand in the usual covariant language: naive uplift of any given vertex from flat space to $AdS$ (by covariantizing all the derivatives) ceases to be gauge invariant due to nonvanishing commutators. A nontrivial statement is that the gauge invariance can be restored by adding terms with less and less derivatives. 

Clearly, one novelty of the cubic deformations of the conformal algebra as compared to those of the Poincare algebra is the $U$-map. On expanding the $U$-map one can see that the vertex has the following general form (say, for $U_L$):
\begin{align}\label{Hexpansion}
    V^{\lambda_1,\lambda_2,\lambda_3}&=U_L V^0= \frac1{z}\sum_{n=0}^{\Lambda-1} \frac{(\PP^L z)^{\Lambda-n}}{\beta_1^{+\lambda_1}\beta_2^{+\lambda_2}\beta_3^{+\lambda_3}}\mathcal{Q}^{\Lambda,n}\,, && \Lambda>0\,.
\end{align}
There is a number of interesting properties: (i) it is polynomial in $z$ and, for this reason, there is no term of order $\PP^0$, the expansion stops exactly at $\PP^1$; (ii) the most important ingredients  $\mathcal{Q}^{\Lambda,n}$, which can be called Metsaev polynomials, are polynomials in $\beta's$ that depend on four combinations thereof: $V_I=\{ \MM, \delbeta, \breb, \MMc=\beta \mathcal{M}\}$;\footnote{More notation: $\beta=\beta_1\beta_2\beta_3$, $\breb=\breb_1\breb_2\breb_3$, $\mathcal{M}=\sum \lambda_i/\beta_i$.} (iii) $\mathcal{Q}^{\Lambda,n}$ has degree $n$ in $\beta$'s, $\mathcal{Q}^{\Lambda,0}=1$; (iv) coefficients of $\mathcal{Q}^{\Lambda,n}$ depend only on $\Lambda$ and not on $\lambda_{1,2,3}$ individually. The last property implies that all vertices with the same total helicity have exactly the same form (up to the overall $\beta$'s). 


\section{Bootstrapping Three-Point Functions}
\label{sec:bootstrap}
We have all the prerequisites to bootstrap the three-point correlation functions in CFT's of Chern-Simons Matter type, i.e. in the three-dimensional CFT's with infinite number of (almost) conserved higher spin tensors. Since we decided to encode the building blocks of three-point correlators in cubic vertices of the bulk dual fields, we will also learn something about Higher Spin Gravities.  

In order to determine the three-point functions we first discuss propagators and establish the dictionary between the bulk helicity fields $\Phi_\lambda(p;z)$ and the boundary conserved tensors. Next, we study the closure of the conformal algebra, which fixes the form of the cubic deformation. Lastly, the three-point functions are easily computed thanks to the simplicity of the propagators and vertices in the light-cone gauge.

\subsection{Propagators and Boundary Conditions}
In order to compute the holographic three-point functions we need the boundary-to-bulk propagators for various kinds of boundary conditions, which we discuss below. The free equations of motion in the light-cone gauge have a remarkably simple form for a massless field of any spin $s=|\lambda|$: 
\begin{align}
    \square\Phi_\lambda(x;z)= (2\pl_+ \pl_-+\pl_1^2 +\pl_z^2)\Phi_\lambda(x;z)&=0\,, &&z>0\,.
\end{align}
Therefore, the propagators for spinning fields are as easy as those for the scalar field. The position space bulk-to-boundary propagator for scalar field that is dual to a weight-$\Delta$ scalar operator is \cite{Freedman:1998tz}
\begin{align}
    K_\Delta(x-y,z)&=\frac{\Gamma(\Delta)}{\pi^{d/2}\Gamma(\Delta-d/2)}\left(\frac{z}{(x-y)^2+z^2}\right)^\Delta \,.
\end{align}
In the case of interest we have $d=3$ and $\Delta=1,2$ and the Fourier transform is\footnote{$\int d^3x\, e^{ipx} K(x,z)$. Note that we removed the factor $z$ from $K(p;z)$ as suggested by the covariant vs. light-cone dictionary, see section \ref{sec:LCAdS} and footnote \ref{ft:action}.}
\begin{align}
    K_{\Delta=1}(p;z)&= - \frac{1}{|p|}e^{-z|p|}\,, &
    K_{\Delta=2}(p;z)&= e^{-z|p|}\,.
\end{align}
As is discussed in \cite{Hartman:2006dy,Giombi:2011ya}, there is a simple relation between the two propagators (and another one between the bulk-to-bulk propagators),
\begin{align}
    K_{\Delta=2}(p;z)=-|p|K_{\Delta=1}(p;z)\,,
\end{align}
which establishes a link between the CFT duals of the same bulk theory with two different boundary conditions, see also \cite{Bekaert:2012ux}. This is one more advantage of the momentum space approach: changing boundary conditions costs an overall factor only (for three-point functions) that does not participate in bulk integration. It is convenient to define
\begin{align}\label{CDelta}
    C_{\Delta=1}&= -{|p|}^{-1}\,, & C_{\Delta=2}&=1\,,
\end{align}
while the boundary-to-bulk propagator is $K_{\Delta}=C_\Delta K$, $K=e^{-z|p|}$.

\subsection{Dictionary}
On $AdS_4$ side we have the bulk helicity fields $\Phi_{\pm s}(p;z)$. The CFT counterpart of imposing the light-cone gauge in $AdS$ is to solve the conservation law $\pl^m J_{ma_2...a_s}=0$ in terms of two functions $\JJ_{\pm  s}(p)$, which may be called {\it light-cone primary operators}, that have definite helicity. Therefore, on the CFT side we have operators $\JJ_{\pm s}(p)$ that are helicity eigen-states. The bulk-to-boundary propagators serve as intertwining operators that map one realization into the other, see e.g. \cite{Metsaev:1999ui,Metsaev:2013kaa,Metsaev:2018xip}. To establish the dictionary we should take the $so(3,2)$-algebra generators $T^{\aAb \aBb}=-T^{ \aBb\aAb}$ in the $AdS_4$-realization, \eqref{helicitybaseAdS}, drag them through the boundary-to-bulk propagator $K$ as to read off the action on the boundary data $O_\lambda(p)$. This action should coincide with the action of $so(3,2)$-generators $T^{\aAb \aBb}$ in the CFT base:
\begin{align}
    T^{\aAb \aBb}_{\text{AdS}} (\Phi_\lambda =KO_\lambda(p))&= K\left(T^{\aAb \aBb}_{\text{CFT}}O_\lambda(p)\right)\,.
\end{align}
This way, the conformal algebra generators on the boundary can be found to be
{\begin{allowdisplaybreaks}
\besubeqs\label{helicitybaseCFT}
\begin{align}
    P^a&= p^a\,,\\
    J^{+1}&= p^+ \pl_p^1- p^1 \pl_p^+\,,\\
    J^{+-}&= p^+\pl_p^- -p^- \pl_p^+\,,\\
    D&= -(2+p^1 \pl_p^1 +p^-\pl_p^++p^+\pl_p^-)\,,\\
    K^+&=-\frac12 p^+ \Delta_p -D\pl_p^+\,,\\
    K^1&=-\frac12 p^1 \Delta_p -D\pl_{p}^1+\lambda \frac{|p|}{p^+}\pl_p^++\frac{\lambda}{|p|}\,,\\
    K^-&= -\frac12 p^-\Delta_p-D\pl_p^--\lambda\frac{|p|}{p^+}\pl_p^1-\frac{\lambda}{|p|}\frac{p^1}{p^+}-\frac{\lambda^2}{\beta}\,,\\
    J^{-1}&=p^- \pl_p^1-p^1 \pl_p^- +\lambda \frac{|p|}{p^+}\,,
\end{align}
\esubeqs
\end{allowdisplaybreaks}}
\noindent where $\Delta_p=2 \pl_p^- \pl_p^+ +\pl_p^1 \pl_p^1$. Eq. \eqref{helicitybaseCFT} is the action on the sources $A_{\pm s}(p)$ for the boundary operators $\JJ_{\pm  s}(p)$. It is the correlators of $\JJ_{\pm s}$ that will be computed.

\subsection{Nonlinear Realization, Higher Spin Gravities}
\label{sec:fucktheshitup}
Section \ref{sec:cubicdeformations} or \cite{Metsaev:2018xip} gives all possible cubic deformations of the conformal algebra $so(3,2)$. We need to fix the one that makes the algebra close at the next order, i.e. quartic in the fields/operators. Surprisingly, this can be done almost without any computation based on the results that are already available, but in a different context. It follows from the structure of the anti-de Sitter algebra $so(3,2)$ that the only relation that needs to be checked is
\begin{align}
    [\JJJ,\HHH]&=0 &&\Longleftrightarrow && [\JJJ_3+\XXX\{\HHH_3\},\HHH_3]=0 \label{HJ}\,.
\end{align}
Indeed, $[\JJJ,\KKK]=0$ and $[\HHH,\KKK]=0$ will follow automatically once \eqref{HJ} is true for $\HHH$ and $\JJJ$ that satisfy the kinematical relations, which is assumed. Based on the structure of $\HHH$ and $\JJJ$, it is not hard to see that the leading terms (highest power in $z$) in the commutator should give exactly the same equations for $C^{\lambda_1,\lambda_2,\lambda_3}$ as in flat space. We see that every Hermitian vertex splits into holomorphic $V$, made out of $\PP_L$ and with $\Lambda>0$, and anti-holomorphic $\bar{V}$ vertices, made out of $\PP_R$ and with $\Lambda<0$. A remarkable property observed in \cite{Metsaev:1991nb,Metsaev:1991mt} is that the closure of the Poincare algebra at the quartic order leads to terms of type $VV$, $\bar{V}\bar{V}$ and $V\bar{V}$, of which the first two receive no contribution from the quartic generators. In other words, there are two sets of equations for the cubic vertices/three-point correlators that must be satisfied irrespective of any contributions from quartic  generators. Likewise, in the $so(3,2)$ case for the holomorphic vertices we find
\begin{align}
\sum_{\omega,\, \text{perm}}\Big[\frac{(\lambda_1+\omega-\lambda_2)\beta_1-(\lambda_2+\omega-\lambda_1)\beta_2}{\beta_1+\beta_2}C^{\lambda_1,\lambda_2,\omega}C^{\lambda_3,\lambda_4,-\omega}\bar{\mathbb{P}}_{12}^{\lambda_1+\lambda_2+\omega-1}\bar{\mathbb{P}}_{34}^{\lambda_3+\lambda_4-\omega}\Big]=0\,,
\label{closure}
\end{align}
where the sum is over all exchanged fields/operators with spin $|\omega|$. Here, $\PP\equiv\PP_L$ and there is a similar equation with $\PP_R$. There is an additional sum over permutations. It is over the four cyclic ones if we are interested in the case when the fields/operators are matrix-valued, i.e. there are some global symmetries on the CFT side, e.g. $U(M)$ or $O(M)$. It is over the six permutations if we consider the minimal case where only even spins are present. It is worth stressing that \eqref{closure} is the crucial equation to bootstrap the Chern-Simons Matter theories and it heavily relies on the specific features of the $3d$ kinematics and on operators being higher spin currents. It would be interesting to derive its analogs for $d>3$ and also to generalize it as to involve contributions of generic 'massive' operators, which should extend the approach to more complicated CFT's with matter in the adjoint of gauge group.

This equation was thoroughly analyzed in the context of the light-front formulation of higher spin theories, \cite{Metsaev:1991nb,Metsaev:1991mt,Ponomarev:2016lrm}. It was shown that it has a unique solution 
\begin{align}\label{Csharp}
    C^{\lambda_1,\lambda_2,\lambda_3}&=\frac{g}{\Gamma[\lambda_1+\lambda_2+\lambda_3]}\,,
\end{align}
where $g$ is a dimensionless coupling constant.\footnote{In the flat space (Poincare algebra) the coupling has to be $g\,l_{Pl}^{\lambda_1+\lambda_2+\lambda_3-1}$ instead of just $g$ by dimensional analysis, where $l_{Pl}$ is the Planck length. In $AdS$ the light-front generators do not depend on the cosmological constant and the appropriate dimension of the vertices is thanks to the $z$-factors. Therefore, there is no need in dimensionful constants for the conformal algebra case. This is a very convenient feature of the light-front approach since there are no dimensionful quantities in CFT's.} The same solution applies to the anti-holomorphic vertices with $\PP_R$ after replacing $\lambda_i\rightarrow -\lambda_i$. 

It is important to discuss the assumptions that lead to \eqref{Csharp}. One assumes at least one spin-$s$ field/operator in the spectrum. Furthermore, the deformation of the algebra is induced by adding $(+s,+s,-s)$ self-interaction vertex. This vertex should lead to nontrivial Ward identities on the CFT side, contrary to the $(+s,+s,+s)$-vertex that gives identically conserved currents. This is the minimal assumption targeting Chern-Simons Matter Theories. Then, it can be seen that \eqref{closure} forces to introduce fields/operators of all spins, including the scalar and graviton/stress-tensor and the solution is \eqref{Csharp}.

Thanks to the holomorphic split described above, one can construct the Chiral Higher Spin Gravities in flat space \cite{Metsaev:1991nb,Metsaev:1991mt,Ponomarev:2016lrm} by taking solution \eqref{Csharp} for $V$ and setting $\bar{V}=0$ (or other way round). Then, the $V\bar{V}$-equations trivialize and we have a consistent theory. Coming back to $AdS_4$, it was already stressed in \cite{Ponomarev:2016lrm,Metsaev:2018xip} that the Chiral Higher Spin Gravity in flat space shares a (overwhelming) number of features with the hypothetical Higher Spin Gravities in $AdS_4$ and, therefore, should admit and $AdS$ uplift. Indeed, we find the same two solutions in $AdS_4/CFT_3$ as well: \eqref{Csharp} and \eqref{Csharp} with $\lambda_i\rightarrow-\lambda_i$. However, these two solutions are non-unitary. 

In the unitary solution, which should be some sort of $V+\bar V$,  we ought to find one more free parameter, which is puzzling in view of uniqueness of \eqref{Csharp}. This follows from the very existence of Chern-Simons Matter Theories and from the analysis of the higher spin algebra cohomology \cite{Sharapov:2018kjz}. The full unitary solution can be obtained via electromagnetic duality transformations. These transformations cannot be realized in terms of the Lorentz covariant bulk fields, but correspond to simple phase rotations in the light-cone gauge, $\Phi_{\pm s}\rightarrow e^{\pm i\theta}\Phi_{\pm s}$. Therefore, what looks like a change of variables in the light-cone gauge results in a completely new action in the covariant formulation. Upon rescaling $g$ appropriately, the full Hermitian cubic action turns out to be
\begin{align}\label{cubicAction}
\begin{aligned}
S=S_2+&g\left( e^{+2i\theta}\VV^{+++}+e^{+i\theta}\VV^{++0}+(\VV^{++-}+\VV^{+00})+e^{-i\theta}\VV^{+-0}+e^{-2i\theta} \VV^{+--}\right)+\\
    &g\left( e^{-2i\theta} \bar\VV^{---}+e^{-i\theta}\bar\VV^{--0}+(\bar\VV^{--+}+\bar\VV^{-00})+e^{+i\theta}\bar\VV^{+-0}+e^{+2i\theta} \bar\VV^{--+}\right)
\,,
\end{aligned}
\end{align}
where $S_2$ is the free part \eqref{freeAction}, $g\sim 1/\tilde N$. We have two phenomenological parameters $\tilde N$ and $\theta$. Here, $\VV$ ($\bar\VV$) is the product of the base interaction vertex $V^{\lambda_1,\lambda_2,\lambda_3}$ and $\Gamma(\Lambda)^{-1}$ or $\Gamma(-\Lambda)^{-1}$, depending on the sign of $\Lambda\equiv \lambda_1+\lambda_2+\lambda_3$. The pluses, minuses or zeros denote the type of the $(\lambda_1, \lambda_2, \lambda_3)$-triplet in the obvious way. Upon setting $\theta=0$ and dropping either the first or the second line in \eqref{cubicAction} we get the cubic action of the (anti)-chiral Higher Spin Theories.

Note that the Einstein-Hilbert cubic vertex is $V^{+2,+2,-2}+V^{-2,-2,+2}$ and corresponds to the phase-independent part of the action, as it should be. In the graviton subsector we find a combination of the Einstein-Hilbert action, Weyl cubed term $C^3$, which is parity even, and the difference between the cubes of the (anti)-self dual components of the Weyl tensor. Schematically, it reads
\begin{align}
    S_{\text{HSGRA}}&=\int \sqrt{g}( R + \cos 2\theta (C_+^3+C_-^3)+\sin 2\theta (C_+^3-C_-^3))+...\,,
\end{align}
where $C^3=C_+^3+C_-^3$. The parity-odd structure already has the right form \cite{Giombi:2011rz}, while both $R$ and $C^3$ give linear combinations of the free fermion and free boson structures, so that the right form of the three point functions is recovered. In general, directly from the action, we see that the holographic correlation functions have the form
\begin{align}
    \langle J_{s_1}J_{s_2}J_{s_3}\rangle&=\tilde N \left(\cos^2\theta \langle J_{s_1}J_{s_2}J_{s_3}\rangle_{F.B.}+\sin\theta\cos\theta\langle J_{s_1}J_{s_2}J_{s_3}\rangle_{Odd}+\sin^2\theta \langle J_{s_1}J_{s_2}J_{s_3}\rangle_{F.F}\right)\,,\notag\\
    \langle J_{s_1}J_{s_2}J_{0}\rangle&=\tilde N \left(\cos\theta \langle J_{s_1}J_{s_2}J_{0}\rangle_{F.B.}+\sin\theta\langle J_{s_1}J_{s_2}J_{0}\rangle_{Odd}\right)\,, \\
    \langle J_{s_1}J_{0}J_{0}\rangle&=\tilde N  \langle J_{s_1}J_{0}J_{0}\rangle_{F.B.}=\tilde N  \langle J_{s_1}J_{0}J_{0}\rangle_{F.F.}\,,\notag
\end{align}
where we chose the $\Delta=1$ boundary conditions and hence labels $F.B.$, $F.F.$. The identification of the odd structure is unambiguous. For $\Delta=2$ boundary conditions we need to swap $F.F.$ and $F.B.$ labels. The $F.F.$ and $F.B.$ contributions can be identified from the free CFT limits. For $\Delta=1$ the odd structure in $\langle J_{s_1}J_{s_2}J_{0}\rangle$ corresponds to that in the Gross-Neveu model. For $\Delta=2$, it corresponds to the one in the critical vector model. There is a unique odd structure for $\langle J_{s_1}J_{s_2}J_{s_3}\rangle$.

To sum up, we have found three solutions for the cubic deformation of the conformal algebra generators. These solutions necessarily contain fields/operators of all spins (at least even). Two of these solutions correspond to the $AdS_4$-uplift of the Chiral Higher Spin Gravities and are non-unitary, but are still interesting both as simple models and as building blocks of the unitary theory. The third solution is unitary and contains one additional parameter that controls parity-breaking, to be associated with the Chern-Simons level $k$. Lastly, \eqref{cubicAction} is a complete cubic action of the $AdS_4$ Higher Spin Gravity, which extends the Type-A results of \cite{Bekaert:2014cea,Sleight:2016dba} and Type-B of \cite{Skvortsov:2015pea}. 

Let us note in passing that, based on the quantum finiteness of the chiral theories in flat space \cite{Skvortsov:2018jea}, we can immediately argue that they should not suffer from UV divergences in the interior of $AdS_4$ as well. The properties of the chiral theories are very close to self-dual Yang-Mills \cite{Chalmers:1996rq} and self-dual Gravity \cite{Krasnov:2016emc}.\footnote{There is a close relation between chiral higher spin gravities in flat space and self-duality \cite{Ponomarev:2017nrr}. Therefore, we can call them either Chiral or Self-Dual Higher Spin Gravities.} Lastly, on changing the signature from $so(3,2)$ to $so(4,1)$ one can extend the chiral theories to de Sitter space. 
 
\subsection{Correlators and Bosonization}
\label{sec:corrandbos}
With all preparations having been made above, the computation of the three-point functions is a one-line exercise --- the momenta do not participate in integration and the $z$-integral is the definition of the $\Gamma$-function. The only care is needed to properly take into account the $U$-map. It is convenient to introduce sources $A_\lambda(p)$ on the boundary. Then, there is an overall factor of 
\begin{align}
    \delta^2(p_1+p_2+p_3)\, C_{\Delta_1}^{\lambda_1}C_{\Delta_2}^{\lambda_2}C_{\Delta_3}^{\lambda_3}\Tr[A_{\lambda_1}A_{\lambda_2}A_{\lambda_3}]
\end{align}
that takes into account possible global symmetries, so that the sources $A_\lambda$ can be matrix-valued. The difference between Dirichlet and Neumann boundary conditions is in $C_\Delta^\lambda$, \eqref{CDelta}, which we can make spin-dependent and mixed boundary conditions are also easy to take into account. The former would correspond to turning on double-trace deformations for various higher spin currents, while the latter to adding Chern-Simons terms, \cite{Giombi:2013yva}. In the simplest case we keep $C^\lambda_\Delta=C_{\Delta=1}$ or can change it to $\Delta=2$ for the scalar, $\lambda=0$, only. The bulk integral is simply\footnote{There is an important issue with the $0-0-0$ correlation function. The $\Delta=1$ bulk integral is divergent, which forces to set the coupling constant to zero (indeed, $1/\Gamma(0)=0$). This implies that the $0-0-0$ correlator for $\Delta=2$ boundary conditions vanishes, in accordance with the critical vector-model. However, in order to reproduce the nonzero $0-0-0$ correlator for $\Delta=1$ boundary conditions one either has to include boundary terms \cite{Freedman:2016yue} or approach $d=3$ from $d>3$ \cite{Bekaert:2014cea}. An alternative suggested by the present approach is to continue in helicity $\Lambda$, $\lim_{\Lambda\rightarrow0}\Gamma(\Lambda)^{-1}\int dz\, e^{-zP} z^{\Lambda-1}=1$. Therefore, we correctly reproduce all the three-point functions, both for $\Delta=1$ and $\Delta=2$ boundary conditions (it is important to take $C_{\Delta}$ into account). } 
\begin{align}
\begin{aligned}
   \int_{AdS_4} V^{\lambda_1,\lambda_2,\lambda_3} =\int dz\, e^{-z(p_1+p_2+p_3)} \sum_{n=0}^{|\Lambda|-1} \mathcal{Q}^{\Lambda}_n\,\PP_{L,R}^{|\Lambda|-n} z^{|\Lambda|-n-1} \prod_a \delta(z_a-z)=\\
    =\sum_{n=0}^{|\Lambda|-1}\mathcal{Q}^{\Lambda}_n\, \Gamma(|\Lambda|-n)\, \left(\frac{1}{3\sqrt{2}P}\left(\sum_a\breb_a(|p_a|\pm p^1_a) \right)\right)^{|\Lambda|-n}\equiv \langle \JJ_{\lambda_1} \JJ_{\lambda_2} \JJ_{\lambda_3}\rangle\,,
\end{aligned}
\end{align}
where $\pm$ corresponds to $\PP_L$ and $\PP_R$ type of vertices and $P=\sum p_i$. Recall that $\mathcal{Q}^{\Lambda}_n$ are Metsaev polynomials from \eqref{Hexpansion}. Therefore, for $+$, i.e. $\PP_L$, we have $\Lambda>0$ and for $-$, i.e. $\PP_R$, we have $\Lambda<0$. Here, $\langle \JJ_{\lambda_1} \JJ_{\lambda_2} \JJ_{\lambda_3}\rangle$ just encodes all possible independent three-point correlation functions (modulo $C_\Delta$). The nontrivial dynamical information is hidden in $C^{\lambda_1,\lambda_2,\lambda_2}$ that stays in front of it in the full cubic vertex $\VV$. The leading singularity of the CFT structure for given helicities $\lambda_{1,2,3}$ is $(p_1+p_2+p_3)^{-|\Lambda|}$.

Let us note that the $AdS_4$ classification of vertices is in one-to-one with that of in $4d$ flat space \cite{Metsaev:2018xip}. The latter is in one-to-one with the spinor-helicity three-point amplitudes, \cite{Benincasa:2007xk,Conde:2016izb}. Lastly, each $AdS_4$ vertex leads to a possible contribution to the three-point $CFT_3$ correlation function. Therefore, $CFT_3$ correlation functions are in one-to-one with the three-point amplitudes in $4d$ flat space, see also \cite{Maldacena:2011nz} for an earlier and \cite{Nagaraj:2018nxq} for the latest discussion.

Also note that for every triplet of spins $s_i=|\lambda_i|$ there can be up to eight independent three-point structures $\langle \JJ_{\lambda_1} \JJ_{\lambda_2} \JJ_{\lambda_3}\rangle$ corresponding to various choices from $+++$ to $---$, whenever possible. On the other hand \cite{Giombi:2011rz,Maldacena:2012sf,Giombi:2016zwa}, the three-point correlation functions of conserved tensors admit three independent structures, two parity-even coming from free boson and free fermion CFT's and an odd one. The resolution of the puzzle is in the unitarity and manifest Lorentz invariance. For example, the Yang-Mills or the Einstein-Hilbert cubic vertex consists of two independent parts in the light-cone gauge $v=V^{+s,+s,-s}$, $\bar{v}=V^{-s,-s,+s}$ for $s=1,2$, respectively. The CPT-invariance requires to take the sum of these two with equal and real coupling constants, i.e. $g(v+\bar v)$ for $g$ real. On the contrary, the all-plus or all-minus vertices $v=V^{+s,+s,+s}$, $\bar v=V^{-s,-s,-s}$ can be added up with complex conjugate couplings $|g|v e^{i\theta}+|g|\bar{v} e^{-i\theta}$ as to produce cubes of the (anti-)selfdual Maxwell tensor, $F_+^3$, $F_-^3$ and Weyl tensor $C_+^3$,  $C_-^3$ for $s=1,2$, respectively. Note that each term separately, $v$ or $\bar{v}$, cannot appear in a unitary theory since the Hamiltonian is not Hermitian. Generalization to correlation functions with higher spin operators is obvious. Undercounting of admissible (but possibly non-unitary) structures within the manifestly covariant approaches also happens for the closely related problem of cubic vertices of massless fields in the $4d$ flat space \cite{Conde:2016izb}.

In particular, the chiral Higher Spin Gravities correspond to the structures only with $\lambda_1+\lambda_2+\lambda_3\equiv\Lambda>0$ or only with $\Lambda<0$, while Hermiticity requires both the sectors. It would be interesting to understand the microscopical realization of the chiral solutions. The flat space studies \cite{Metsaev:1991nb,Metsaev:1991mt,Ponomarev:2016lrm,Ponomarev:2017nrr,Skvortsov:2018jea} show that the chiral theories are quite special in that they are not obstructed by the usual non-locality problems of higher spin theories. Given the results of \cite{Bekaert:2015tva,Sleight:2017pcz,Ponomarev:2017qab}, it well may be that the chiral theories are the only higher spin theories that can be given a bulk definition that is independent of their CFT dual.\footnote{It is worth mentioning that the Vasiliev equations \cite{Vasiliev:1990cm} suffer from even more severe non-localities \cite{Boulanger:2015ova,Skvortsov:2015lja} than HSGRA are known to have \cite{Bekaert:2015tva,Sleight:2017pcz,Ponomarev:2017qab}.  Nevertheless, they are based on the right symmetry algebra and have the correct spectrum of fields. There are attempts to fix the problem \cite{Vasiliev:2016xui} with some positive results \cite{Misuna:2017bjb} that can be compared to ours. }

The spectrum of the chiral theories is the usual one --- massless fields of all spins. It is the interactions that violate unitarity in a way that is similar to self-dual Yang-Mills and self-dual Gravity. Therefore, the CFT dual is expected to be of Chern-Simons Matter type, but with some non-unitary limit taken. The fact that we have two chiral limits is reminiscent of the fishnet theories \cite{Gurdogan:2015csr,Caetano:2016ydc}. The fishnet theories result from a certain double-scaling limit of $\gamma$-twisted $\mathcal{N}=4$ SYM or ABJ(M) theories. In the gravitational dual description the limit corresponds to $\lambda = R_{AdS}^2/l_s^2$ taken to zero. This is exactly the tensionless limit where higher spin theories are expected to emerge \cite{Sundborg:2000wp,Sezgin:2002rt,Beisert:2004di}. Indeed, the free (or weakly coupled) SYM has (almost) conserved higher spin tensors that are dual to massless higher spin fields in $AdS_5$. Likewise, the fishnet theories have almost conserved higher spin currents and admit a free limit where the currents are identically conserved. It is difficult to say at the moment what happens on the gravitational side, but, taking into account the discussion above, we can refer to the fishnet limit as to a tensionless one. Therefore, fishnet theories operate in the same regime as CFT duals of higher spin theories, the important subtlety being that higher spin theories are better understood as duals of weakly coupled (or even free) CFT's with matter in vectorial representations \cite{Sezgin:2002rt,Klebanov:2002ja}. Interestingly enough ABJ theories admit both fishnet limits \cite{Caetano:2016ydc} and vector-like limits \cite{Chang:2012kt}. Therefore, it looks plausible that the chiral theories may emerge in some tensionless limit and are dual to CFT's of the fishnet type that may result from certain twists and limits of the already known Chern-Simons Matter theories. In this regard, it is worth noting that the chiral theories can be super-symmetrized.

Summarizing our findings, the final result for the three-point functions for $3d$ CFT's with higher spin currents can be written in the form of an effective action as
\begin{align}\notag
    W_3[A]&= \sum_{\lambda_a} C_{\Delta_1}^{\lambda_1}C_{\Delta_2}^{\lambda_2}C_{\Delta_3}^{\lambda_3} C^{\lambda_1,\lambda_2,\lambda_3}\int \prod_i d^3p_i\,  \delta^3 (p_1+p_2+p_3)\Tr[A_{\lambda_1}A_{\lambda_2}A_{\lambda_3}]\langle \JJ_{\lambda_1} \JJ_{\lambda_2} \JJ_{\lambda_3}\rangle \,,
\end{align}
where the structure constants $C^{\lambda_1,\lambda_2,\lambda_3}$ correspond to one of the three solutions found in section \ref{sec:fucktheshitup}: chiral, anti-chiral and the unitary one. The latter gives all three-point functions of single-trace operators in Chern-Simons Matter theories without having to use any microscopical realization of these theories. In particular, we do not have to make any assumptions on whether the fundamental constituents are bosons or fermions. This manifests the bosonization duality to this order. It would be interesting to extend the analysis of the nonlinear realization to higher orders, of course. An obvious complication is that the quartic analog of \eqref{closure} should contain a unknown function of two variables (analogs of conformal cross-ratios). Nevertheless, a considerable part of the four-point function should be given by the chiral solutions, which is an immediate advantage that does not seem to be available in other approaches. 

\section*{Acknowledgments}
\label{sec:Aknowledgements}
We are grateful to  Karapet Mkrtchyan, Dmitry Ponomarev, Stefan Theisen, Arkady Tseytlin, Sasha Zhiboedov and especially to Ruslan Metsaev for useful discussions. 
The work of E.S. was supported by the Russian Science Foundation grant 18-72-10123 in association with the Lebedev Physical Institute. 

\begin{appendix}
\renewcommand{\thesection}{\Alph{section}}
\renewcommand{\theequation}{\Alph{section}.\arabic{equation}}
\setcounter{equation}{0}\setcounter{section}{0}

\end{appendix}

\setstretch{1.0}
\providecommand{\href}[2]{#2}\begingroup\raggedright\endgroup


\begin{thebibliography}{10}

\bibitem{Goddard:1973qh}
P.~Goddard, J.~Goldstone, C.~Rebbi and C.~B. Thorn, {\it {Quantum dynamics of a
  massless relativistic string}},  {\em Nucl. Phys.} {\bf B56} (1973) 109--135.

\bibitem{Bengtsson:1983pg}
A.~K.~H. Bengtsson, I.~Bengtsson and L.~Brink, {\it {Cubic Interaction Terms
  for Arbitrarily Extended Supermultiplets}},  {\em Nucl. Phys.} {\bf B227}
  (1983) 41--49.

\bibitem{Metsaev:1991nb}
R.~R. Metsaev, {\it {S matrix approach to massless higher spins theory. 2: The
  Case of internal symmetry}},  {\em Mod. Phys. Lett.} {\bf A6} (1991)
  2411--2421.

\bibitem{Brink:2005wh}
L.~Brink, {\it {Particle physics as representations of the Poincare algebra}},
  in {\em {Proceedings, Symposium Henri Poincare, Brussels, Belgium, 8-9 Oct
  2004}}, p.~125, 2005.
\newblock \href{http://arXiv.org/abs/hep-th/0503035}{{\tt hep-th/0503035}}.

\bibitem{Metsaev:2018xip}
R.~R. Metsaev, {\it {Light-cone gauge cubic interaction vertices for massless
  fields in AdS(4)}},  \href{http://arXiv.org/abs/1807.07542}{{\tt
  1807.07542}}.

\bibitem{Giombi:2011kc}
S.~Giombi, S.~Minwalla, S.~Prakash, S.~P. Trivedi, S.~R. Wadia and X.~Yin, {\it
  {Chern-Simons Theory with Vector Fermion Matter}},  {\em Eur. Phys. J.} {\bf
  C72} (2012) 2112 [\href{http://arXiv.org/abs/1110.4386}{{\tt 1110.4386}}].

\bibitem{Maldacena:2012sf}
J.~Maldacena and A.~Zhiboedov, {\it {Constraining conformal field theories with
  a slightly broken higher spin symmetry}},  {\em Class. Quant. Grav.} {\bf 30}
  (2013) 104003 [\href{http://arXiv.org/abs/1204.3882}{{\tt 1204.3882}}].

\bibitem{Aharony:2012nh}
O.~Aharony, G.~Gur-Ari and R.~Yacoby, {\it {Correlation Functions of Large N
  Chern-Simons-Matter Theories and Bosonization in Three Dimensions}},  {\em
  JHEP} {\bf 12} (2012) 028 [\href{http://arXiv.org/abs/1207.4593}{{\tt
  1207.4593}}].

\bibitem{GurAri:2012is}
G.~Gur-Ari and R.~Yacoby, {\it {Correlators of Large N Fermionic Chern-Simons
  Vector Models}},  {\em JHEP} {\bf 02} (2013) 150
  [\href{http://arXiv.org/abs/1211.1866}{{\tt 1211.1866}}].

\bibitem{Aharony:2015mjs}
O.~Aharony, {\it {Baryons, monopoles and dualities in Chern-Simons-matter
  theories}},  {\em JHEP} {\bf 02} (2016) 093
  [\href{http://arXiv.org/abs/1512.00161}{{\tt 1512.00161}}].

\bibitem{Karch:2016sxi}
A.~Karch and D.~Tong, {\it {Particle-Vortex Duality from 3d Bosonization}},
  {\em Phys. Rev.} {\bf X6} (2016), no.~3 031043
  [\href{http://arXiv.org/abs/1606.01893}{{\tt 1606.01893}}].

\bibitem{Seiberg:2016gmd}
N.~Seiberg, T.~Senthil, C.~Wang and E.~Witten, {\it {A Duality Web in 2+1
  Dimensions and Condensed Matter Physics}},  {\em Annals Phys.} {\bf 374}
  (2016) 395--433 [\href{http://arXiv.org/abs/1606.01989}{{\tt 1606.01989}}].

\bibitem{Gurdogan:2015csr}
{\"O}.~G{\"u}rdo{\u g}an and V.~Kazakov, {\it {New Integrable 4D Quantum Field
  Theories from Strongly Deformed Planar $\mathcal N = $ 4 Supersymmetric
  Yang-Mills Theory}},  {\em Phys. Rev. Lett.} {\bf 117} (2016), no.~20 201602
  [\href{http://arXiv.org/abs/1512.06704}{{\tt 1512.06704}}]. [Addendum: Phys.
  Rev. Lett.117,no.25,259903(2016)].

\bibitem{Caetano:2016ydc}
J.~Caetano, {\"O}.~G{\"u}rdo{\u g}an and V.~Kazakov, {\it {Chiral limit of $
  \mathcal{N} $ = 4 SYM and ABJM and integrable Feynman graphs}},  {\em JHEP}
  {\bf 03} (2018) 077 [\href{http://arXiv.org/abs/1612.05895}{{\tt
  1612.05895}}].

\bibitem{Maldacena:1997re}
J.~M. Maldacena, {\it {The large N limit of superconformal field theories and
  supergravity}},  {\em Adv. Theor. Math. Phys.} {\bf 2} (1998) 231--252
  [\href{http://arXiv.org/abs/hep-th/9711200}{{\tt hep-th/9711200}}].

\bibitem{Witten:1998qj}
E.~Witten, {\it {Anti-de Sitter space and holography}},  {\em Adv. Theor. Math.
  Phys.} {\bf 2} (1998) 253--291
  [\href{http://arXiv.org/abs/hep-th/9802150}{{\tt hep-th/9802150}}].

\bibitem{Gubser:1998bc}
S.~S. Gubser, I.~R. Klebanov and A.~M. Polyakov, {\it {Gauge theory correlators
  from non-critical string theory}},  {\em Phys. Lett.} {\bf B428} (1998)
  105--114 [\href{http://arXiv.org/abs/hep-th/9802109}{{\tt hep-th/9802109}}].

\bibitem{Klebanov:2002ja}
I.~R. Klebanov and A.~M. Polyakov, {\it {AdS dual of the critical O(N) vector
  model}},  {\em Phys. Lett.} {\bf B550} (2002) 213--219
  [\href{http://arXiv.org/abs/hep-th/0210114}{{\tt hep-th/0210114}}].

\bibitem{Sezgin:2002rt}
E.~Sezgin and P.~Sundell, {\it {Massless higher spins and holography}},  {\em
  Nucl.Phys.} {\bf B644} (2002) 303--370
  [\href{http://arXiv.org/abs/hep-th/0205131}{{\tt hep-th/0205131}}].

\bibitem{Sezgin:2003pt}
E.~Sezgin and P.~Sundell, {\it {Holography in 4D (super) higher spin theories
  and a test via cubic scalar couplings}},  {\em JHEP} {\bf 0507} (2005) 044
  [\href{http://arXiv.org/abs/hep-th/0305040}{{\tt hep-th/0305040}}].

\bibitem{Leigh:2003gk}
R.~G. Leigh and A.~C. Petkou, {\it {Holography of the N=1 higher spin theory on
  AdS(4)}},  {\em JHEP} {\bf 0306} (2003) 011
  [\href{http://arXiv.org/abs/hep-th/0304217}{{\tt hep-th/0304217}}].

\bibitem{Deser:1980fk}
S.~Deser and H.~Nicolai, {\it {Nonabelian Zilch}},  {\em Phys. Lett.} {\bf 98B}
  (1981) 45--47.

\bibitem{Maldacena:2011jn}
J.~Maldacena and A.~Zhiboedov, {\it {Constraining Conformal Field Theories with
  A Higher Spin Symmetry}},  \href{http://arXiv.org/abs/1112.1016}{{\tt
  1112.1016}}.

\bibitem{Boulanger:2013zza}
N.~Boulanger, D.~Ponomarev, E.~Skvortsov and M.~Taronna, {\it {On the
  uniqueness of higher-spin symmetries in AdS and CFT}},
  \href{http://arXiv.org/abs/1305.5180}{{\tt 1305.5180}}.

\bibitem{Alba:2013yda}
V.~Alba and K.~Diab, {\it {Constraining conformal field theories with a higher
  spin symmetry in d=4}},  \href{http://arXiv.org/abs/1307.8092}{{\tt
  1307.8092}}.

\bibitem{Alba:2015upa}
V.~Alba and K.~Diab, {\it {Constraining conformal field theories with a higher
  spin symmetry in $d> 3$ dimensions}},
  \href{http://arXiv.org/abs/1510.02535}{{\tt 1510.02535}}.

\bibitem{Aharony:2012ns}
O.~Aharony, S.~Giombi, G.~Gur-Ari, J.~Maldacena and R.~Yacoby, {\it {The
  Thermal Free Energy in Large N Chern-Simons-Matter Theories}},  {\em JHEP}
  {\bf 03} (2013) 121 [\href{http://arXiv.org/abs/1211.4843}{{\tt 1211.4843}}].

\bibitem{Jain:2013py}
S.~Jain, S.~Minwalla, T.~Sharma, T.~Takimi, S.~R. Wadia and S.~Yokoyama, {\it
  {Phases of large $N$ vector Chern-Simons theories on $S^2 \times S^1$}},
  {\em JHEP} {\bf 09} (2013) 009 [\href{http://arXiv.org/abs/1301.6169}{{\tt
  1301.6169}}].

\bibitem{Giombi:2016zwa}
S.~Giombi, V.~Gurucharan, V.~Kirilin, S.~Prakash and E.~Skvortsov, {\it {On the
  Higher-Spin Spectrum in Large N Chern-Simons Vector Models}},  {\em JHEP}
  {\bf 01} (2017) 058 [\href{http://arXiv.org/abs/1610.08472}{{\tt
  1610.08472}}].

\bibitem{ElShowk:2012ht}
S.~El-Showk, M.~F. Paulos, D.~Poland, S.~Rychkov, D.~Simmons-Duffin and
  A.~Vichi, {\it {Solving the 3D Ising Model with the Conformal Bootstrap}},
  {\em Phys. Rev.} {\bf D86} (2012) 025022
  [\href{http://arXiv.org/abs/1203.6064}{{\tt 1203.6064}}].

\bibitem{Giombi:2016hkj}
S.~Giombi and V.~Kirilin, {\it {Anomalous dimensions in CFT with weakly broken
  higher spin symmetry}},  {\em JHEP} {\bf 11} (2016) 068
  [\href{http://arXiv.org/abs/1601.01310}{{\tt 1601.01310}}].

\bibitem{Skvortsov:2015pea}
E.~D. Skvortsov, {\it {On (Un)Broken Higher-Spin Symmetry in Vector Models}},
  in {\em {Proceedings, International Workshop on Higher Spin Gauge Theories:
  Singapore, Singapore, November 4-6, 2015}}, pp.~103--137, 2017.
\newblock \href{http://arXiv.org/abs/1512.05994}{{\tt 1512.05994}}.

\bibitem{Aharony:2018npf}
O.~Aharony, L.~F. Alday, A.~Bissi and R.~Yacoby, {\it {The Analytic Bootstrap
  for Large $N$ Chern-Simons Vector Models}},  {\em JHEP} {\bf 08} (2018) 166
  [\href{http://arXiv.org/abs/1805.04377}{{\tt 1805.04377}}].

\bibitem{Alday:2016jfr}
L.~F. Alday, {\it {Solving CFTs with Weakly Broken Higher Spin Symmetry}},
  {\em JHEP} {\bf 10} (2017) 161 [\href{http://arXiv.org/abs/1612.00696}{{\tt
  1612.00696}}].

\bibitem{Charan:2017jyc}
V.~Guru~Charan and S.~Prakash, {\it {On the Higher Spin Spectrum of
  Chern-Simons Theory coupled to Fermions in the Large Flavour Limit}},  {\em
  JHEP} {\bf 02} (2018) 094 [\href{http://arXiv.org/abs/1711.11300}{{\tt
  1711.11300}}].

\bibitem{Yacoby:2018yvy}
R.~Yacoby, {\it {Scalar Correlators in Bosonic Chern-Simons Vector Models}},
  \href{http://arXiv.org/abs/1805.11627}{{\tt 1805.11627}}.

\bibitem{Sleight:2018ryu}
C.~Sleight and M.~Taronna, {\it {A Note on Anomalous Dimensions from Crossing
  Kernels}},  \href{http://arXiv.org/abs/1807.05941}{{\tt 1807.05941}}.

\bibitem{Turiaci:2018dht}
G.~J. Turiaci and A.~Zhiboedov, {\it {Veneziano Amplitude of Vasiliev Theory}},
   {\em JHEP} {\bf 10} (2018) 034 [\href{http://arXiv.org/abs/1802.04390}{{\tt
  1802.04390}}].

\bibitem{Sharapov:2018kjz}
A.~Sharapov and E.~Skvortsov, {\it {$A_\infty$ Algebras from Slightly Broken
  Higher Spin Symmetries}},  \href{http://arXiv.org/abs/1809.10027}{{\tt
  1809.10027}}.

\bibitem{Bekaert:2015tva}
X.~Bekaert, J.~Erdmenger, D.~Ponomarev and C.~Sleight, {\it {Quartic AdS
  Interactions in Higher-Spin Gravity from Conformal Field Theory}},  {\em
  JHEP} {\bf 11} (2015) 149 [\href{http://arXiv.org/abs/1508.04292}{{\tt
  1508.04292}}].

\bibitem{Sleight:2017pcz}
C.~Sleight and M.~Taronna, {\it {Higher-Spin Gauge Theories and Bulk
  Locality}},  {\em Phys. Rev. Lett.} {\bf 121} (2018), no.~17 171604
  [\href{http://arXiv.org/abs/1704.07859}{{\tt 1704.07859}}].

\bibitem{Ponomarev:2017qab}
D.~Ponomarev, {\it {A Note on (Non)-Locality in Holographic Higher Spin
  Theories}},  {\em Universe} {\bf 4} (2018), no.~1 2
  [\href{http://arXiv.org/abs/1710.00403}{{\tt 1710.00403}}].

\bibitem{Heemskerk:2009pn}
I.~Heemskerk, J.~Penedones, J.~Polchinski and J.~Sully, {\it {Holography from
  Conformal Field Theory}},  {\em JHEP} {\bf 10} (2009) 079
  [\href{http://arXiv.org/abs/0907.0151}{{\tt 0907.0151}}].

\bibitem{Sleight:2016dba}
C.~Sleight and M.~Taronna, {\it {Higher Spin Interactions from Conformal Field
  Theory: The Complete Cubic Couplings}},  {\em Phys. Rev. Lett.} {\bf 116}
  (2016), no.~18 181602 [\href{http://arXiv.org/abs/1603.00022}{{\tt
  1603.00022}}].

\bibitem{Koch:2010cy}
R.~de~Mello~Koch, A.~Jevicki, K.~Jin and J.~P. Rodrigues, {\it {$AdS_4/CFT_3$
  Construction from Collective Fields}},  {\em Phys. Rev.} {\bf D83} (2011)
  025006 [\href{http://arXiv.org/abs/1008.0633}{{\tt 1008.0633}}].

\bibitem{Giombi:2011rz}
S.~Giombi, S.~Prakash and X.~Yin, {\it {A Note on CFT Correlators in Three
  Dimensions}},  \href{http://arXiv.org/abs/1104.4317}{{\tt 1104.4317}}.

\bibitem{Colombo:2012jx}
N.~Colombo and P.~Sundell, {\it {Higher Spin Gravity Amplitudes From Zero-form
  Charges}},  \href{http://arXiv.org/abs/1208.3880}{{\tt 1208.3880}}.

\bibitem{Didenko:2012tv}
V.~Didenko and E.~Skvortsov, {\it {Exact higher-spin symmetry in CFT: all
  correlators in unbroken Vasiliev theory}},  {\em JHEP} {\bf 1304} (2013) 158
  [\href{http://arXiv.org/abs/1210.7963}{{\tt 1210.7963}}].

\bibitem{Didenko:2013bj}
V.~E. Didenko, J.~Mei and E.~D. Skvortsov, {\it {Exact higher-spin symmetry in
  CFT: free fermion correlators from Vasiliev Theory}},  {\em Phys. Rev.} {\bf
  D88} (2013) 046011 [\href{http://arXiv.org/abs/1301.4166}{{\tt 1301.4166}}].

\bibitem{Bonezzi:2017vha}
R.~Bonezzi, N.~Boulanger, D.~De~Filippi and P.~Sundell, {\it {Noncommutative
  Wilson lines in higher-spin theory and correlation functions of conserved
  currents for free conformal fields}},  {\em J. Phys.} {\bf A50} (2017),
  no.~47 475401 [\href{http://arXiv.org/abs/1705.03928}{{\tt 1705.03928}}].

\bibitem{Benincasa:2007xk}
P.~Benincasa and F.~Cachazo, {\it {Consistency Conditions on the S-Matrix of
  Massless Particles}},  \href{http://arXiv.org/abs/0705.4305}{{\tt
  0705.4305}}.

\bibitem{Conde:2016izb}
E.~Conde, E.~Joung and K.~Mkrtchyan, {\it {Spinor-Helicity Three-Point
  Amplitudes from Local Cubic Interactions}},  {\em JHEP} {\bf 08} (2016) 040
  [\href{http://arXiv.org/abs/1605.07402}{{\tt 1605.07402}}].

\bibitem{Bengtsson:1983pd}
A.~K.~H. Bengtsson, I.~Bengtsson and L.~Brink, {\it {Cubic Interaction Terms
  for Arbitrary Spin}},  {\em Nucl. Phys.} {\bf B227} (1983) 31--40.

\bibitem{Maldacena:2011nz}
J.~M. Maldacena and G.~L. Pimentel, {\it {On graviton non-Gaussianities during
  inflation}},  {\em JHEP} {\bf 09} (2011) 045
  [\href{http://arXiv.org/abs/1104.2846}{{\tt 1104.2846}}].

\bibitem{Nagaraj:2018nxq}
B.~Nagaraj and D.~Ponomarev, {\it {Spinor-Helicity Formalism for Massless
  Fields in AdS${}_4$}},  \href{http://arXiv.org/abs/1811.08438}{{\tt
  1811.08438}}.

\bibitem{Metsaev:1991mt}
R.~R. Metsaev, {\it {Poincare invariant dynamics of massless higher spins:
  Fourth order analysis on mass shell}},  {\em Mod. Phys. Lett.} {\bf A6}
  (1991) 359--367.

\bibitem{Ponomarev:2016lrm}
D.~Ponomarev and E.~D. Skvortsov, {\it {Light-Front Higher-Spin Theories in
  Flat Space}},  {\em J. Phys.} {\bf A50} (2017), no.~9 095401
  [\href{http://arXiv.org/abs/1609.04655}{{\tt 1609.04655}}].

\bibitem{Ponomarev:2017nrr}
D.~Ponomarev, {\it {Chiral Higher Spin Theories and Self-Duality}},  {\em JHEP}
  {\bf 12} (2017) 141 [\href{http://arXiv.org/abs/1710.00270}{{\tt
  1710.00270}}].

\bibitem{Skvortsov:2018jea}
E.~D. Skvortsov, T.~Tran and M.~Tsulaia, {\it {Quantum Chiral Higher Spin
  Gravity}},  {\em Phys. Rev. Lett.} {\bf 121} (2018), no.~3 031601
  [\href{http://arXiv.org/abs/1805.00048}{{\tt 1805.00048}}].

\bibitem{Kontsevich:1997vb}
M.~Kontsevich, {\it {Deformation quantization of Poisson manifolds. 1.}},  {\em
  Lett. Math. Phys.} {\bf 66} (2003) 157--216
  [\href{http://arXiv.org/abs/q-alg/9709040}{{\tt q-alg/9709040}}].

\bibitem{Metsaev:1999ui}
R.~R. Metsaev, {\it Light cone form of field dynamics in anti-de sitter
  spacetime and ads/cft correspondence},  {\em Nucl. Phys.} {\bf B563} (1999)
  295--348 [\href{http://arXiv.org/abs/hep-th/9906217}{{\tt hep-th/9906217}}].

\bibitem{Metsaev:2013kaa}
R.~R. Metsaev, {\it {Light-cone gauge approach to arbitrary spin fields,
  currents, and shadows}},  {\em J. Phys.} {\bf A47} (2014) 375401
  [\href{http://arXiv.org/abs/1312.5679}{{\tt 1312.5679}}].

\bibitem{Bzowski:2013sza}
A.~Bzowski, P.~McFadden and K.~Skenderis, {\it {Implications of conformal
  invariance in momentum space}},  {\em JHEP} {\bf 03} (2014) 111
  [\href{http://arXiv.org/abs/1304.7760}{{\tt 1304.7760}}].

\bibitem{Mack:1969dg}
G.~Mack and I.~Todorov, {\it {Irreducibility of the ladder representations of
  u(2,2) when restricted to the poincare subgroup}},  {\em J. Math. Phys.} {\bf
  10} (1969) 2078--2085.

\bibitem{Metsaev:1995jp}
R.~R. Metsaev, {\it {All conformal invariant representations of d-dimensional
  anti-de Sitter group}},  {\em Mod. Phys. Lett.} {\bf A10} (1995) 1719--1731.

\bibitem{Freedman:1998tz}
D.~Z. Freedman, S.~D. Mathur, A.~Matusis and L.~Rastelli, {\it {Correlation
  functions in the CFT(d) / AdS(d+1) correspondence}},  {\em Nucl. Phys.} {\bf
  B546} (1999) 96--118 [\href{http://arXiv.org/abs/hep-th/9804058}{{\tt
  hep-th/9804058}}].

\bibitem{Hartman:2006dy}
T.~Hartman and L.~Rastelli, {\it {Double-trace deformations, mixed boundary
  conditions and functional determinants in AdS/CFT}},  {\em JHEP} {\bf 0801}
  (2008) 019 [\href{http://arXiv.org/abs/hep-th/0602106}{{\tt
  hep-th/0602106}}].

\bibitem{Giombi:2011ya}
S.~Giombi and X.~Yin, {\it {On Higher Spin Gauge Theory and the Critical O(N)
  Model}},  {\em Phys.Rev.} {\bf D85} (2012) 086005
  [\href{http://arXiv.org/abs/1105.4011}{{\tt 1105.4011}}].

\bibitem{Bekaert:2012ux}
X.~Bekaert, E.~Joung and J.~Mourad, {\it {Comments on higher-spin holography}},
   \href{http://arXiv.org/abs/1202.0543}{{\tt 1202.0543}}.

\bibitem{Bekaert:2014cea}
X.~Bekaert, J.~Erdmenger, D.~Ponomarev and C.~Sleight, {\it {Towards
  holographic higher-spin interactions: Four-point functions and higher-spin
  exchange}},  {\em JHEP} {\bf 03} (2015) 170
  [\href{http://arXiv.org/abs/1412.0016}{{\tt 1412.0016}}].

\bibitem{Chalmers:1996rq}
G.~Chalmers and W.~Siegel, {\it {The Selfdual sector of QCD amplitudes}},  {\em
  Phys. Rev.} {\bf D54} (1996) 7628--7633
  [\href{http://arXiv.org/abs/hep-th/9606061}{{\tt hep-th/9606061}}].

\bibitem{Krasnov:2016emc}
K.~Krasnov, {\it {Self-Dual Gravity}},  {\em Class. Quant. Grav.} {\bf 34}
  (2017), no.~9 095001 [\href{http://arXiv.org/abs/1610.01457}{{\tt
  1610.01457}}].

\bibitem{Giombi:2013yva}
S.~Giombi, I.~R. Klebanov, S.~S. Pufu, B.~R. Safdi and G.~Tarnopolsky, {\it
  {AdS Description of Induced Higher-Spin Gauge Theory}},  {\em JHEP} {\bf 10}
  (2013) 016 [\href{http://arXiv.org/abs/1306.5242}{{\tt 1306.5242}}].

\bibitem{Freedman:2016yue}
D.~Z. Freedman, K.~Pilch, S.~S. Pufu and N.~P. Warner, {\it {Boundary Terms and
  Three-Point Functions: An AdS/CFT Puzzle Resolved}},  {\em JHEP} {\bf 06}
  (2017) 053 [\href{http://arXiv.org/abs/1611.01888}{{\tt 1611.01888}}].

\bibitem{Vasiliev:1990cm}
M.~A. Vasiliev, {\it {Closed equations for interacting gauge fields of all
  spins}},  {\em JETP Lett.} {\bf 51} (1990) 503--507. [Pisma Zh. Eksp. Teor.
  Fiz.51,446(1990)].

\bibitem{Boulanger:2015ova}
N.~Boulanger, P.~Kessel, E.~D. Skvortsov and M.~Taronna, {\it {Higher spin
  interactions in four-dimensions: Vasiliev versus Fronsdal}},  {\em J. Phys.}
  {\bf A49} (2016), no.~9 095402 [\href{http://arXiv.org/abs/1508.04139}{{\tt
  1508.04139}}].

\bibitem{Skvortsov:2015lja}
E.~D. Skvortsov and M.~Taronna, {\it {On Locality, Holography and Unfolding}},
  {\em JHEP} {\bf 11} (2015) 044 [\href{http://arXiv.org/abs/1508.04764}{{\tt
  1508.04764}}].

\bibitem{Vasiliev:2016xui}
M.~A. Vasiliev, {\it {Current Interactions and Holography from the 0-Form
  Sector of Nonlinear Higher-Spin Equations}},  {\em JHEP} {\bf 10} (2017) 111
  [\href{http://arXiv.org/abs/1605.02662}{{\tt 1605.02662}}].

\bibitem{Misuna:2017bjb}
N.~Misuna, {\it {On current contribution to Fronsdal equations}},  {\em Phys.
  Lett.} {\bf B778} (2018) 71--78 [\href{http://arXiv.org/abs/1706.04605}{{\tt
  1706.04605}}].

\bibitem{Sundborg:2000wp}
B.~Sundborg, {\it {Stringy gravity, interacting tensionless strings and
  massless higher spins}},  {\em Nucl. Phys. Proc. Suppl.} {\bf 102} (2001)
  113--119 [\href{http://arXiv.org/abs/hep-th/0103247}{{\tt hep-th/0103247}}].

\bibitem{Beisert:2004di}
N.~Beisert, M.~Bianchi, J.~F. Morales and H.~Samtleben, {\it {Higher spin
  symmetry and N=4 SYM}},  {\em JHEP} {\bf 07} (2004) 058
  [\href{http://arXiv.org/abs/hep-th/0405057}{{\tt hep-th/0405057}}].

\bibitem{Chang:2012kt}
C.-M. Chang, S.~Minwalla, T.~Sharma and X.~Yin, {\it {ABJ Triality: from Higher
  Spin Fields to Strings}},  \href{http://arXiv.org/abs/1207.4485}{{\tt
  1207.4485}}.

\end{thebibliography}
\end{document}